\documentclass[twocolumn,showpacs,preprintnumbers]{revtex4}
\usepackage{amssymb}
\usepackage{amsfonts}
\usepackage{amsmath}
\usepackage{graphicx}
\usepackage{dcolumn}
\usepackage{bm}

\input{tcilatex}

\begin{document}

\title{The quantum approach to human reasoning does explain the belief-bias
effect}
\author{E. D. Vol}
\email{vol@ilt.kharkov.ua}
\affiliation{B. Verkin Institute for Low Temperature Physics and Engineering of the
National Academy of Sciences of Ukraine 47, Lenin Ave., Kharkov 61103,
Ukraine.}
\date{\today }

\begin{abstract}
Based on the ideas of quantum physics and dual-process theory of human
reasoning that takes into account two primary mechanisms of reasoning : 1)
deductive rational thinking and 2) intuitive heuristic judgment, we proposed
the \ "quantum" approach to practical human logic that allows one  to
specify the most distinctive peculiarities in activity of two reasoning
systems mentioned above and in addition to describe phenomenologically 
well-established experimentally belief-bias effect .
\end{abstract}

\maketitle

\section{Introduction}

The idea that some essential human values and concepts may be incompatable
with each other had originated long before the beginning of scientific
psychology. By distinct ways this idea was justified by such outstanding
philosophers and thinkers as G.W. Leibniz, N.Machiavelli and I. Kant. The
interested reader can find detail account of the history of this idea with
relevant references in \cite{1s}. However only in the XX century with the
rise of quantum theory this idea has received adequate scientific expression
in the language of the Bohr's Complementarity Principle. We shall give here
only two distinctive quotations of founding fathers of quantum mechanics
that are clearly demonstrating their profound understanding of the
inconsistancy of some basic concepts relating to human psychology. So, in
the paper of 1948 "On the Notions of Causality and Complementarity" N. Bohr
wrote:"Recognition of complementary relationship is not least required in
psychology, where the conditions for analysis and synthesis of experience
exhibit striking analogy with the situation in atomic physis. In fact, the
use of words like thoughts, and sentiments, equally indispensable to
illustrate the diversity of psychical experience, pertain to mutually
exclusive situations characterized by a different drawing of the line of
separation between subject and object. In particular, the place left for the
feeling of volition is afforded by the very circumstance that situations
where we experience freedom of will are incompatible with psychological
situations where causal analysis is reasonably attempted. In other words,
when we use the phrase "I will" we renounce explanatory argumentation. In
fact, the use which we make of words like "thought" and "feeling," or
"instinct" and "reason" to describe psychic experiences of different types,
shows the existence of characteristic relationships of complementarity
conditioned by the peculiarity of introspection" \cite{2s}. On the other
hand W.Pauli drew particular attention to the problem of relation between
complementarity of mental and physical aspects of the same reality. In his
inspiring paper "The influence of \ archetypal ideas on the scientific
theories of Kepler" \cite{2s} he wrote:" \textquotedblleft The general
problem of the relationship between psyche and physics, between inside and
outside, will hardly be solved with the notion of a `psychophysical
parallelism, put forward in the past century. However, modern science has
perhaps brought us closer to a more satisfying conception of this
relationship insofar as it introduced the concept of complementarity within
physics. It would be most satisfactory if physis and psyche could be
conceived as complementary aspects of the same reality.\textquotedblright
.Unfortunately at the time these deep ideas are not influenced the
development of experimental psychology. All the more remarkable that modern
cognitive psychology irrespectively came in fact to the similar
conclusions.In particular numerous experts in so different areas of
cognitive psychology as attention,memory,decision making, learning with one
accord believe that dual processes and dual systems play fundamental role
for nearly all basic cognitive mechanisms in human mind (see e.g. \cite{4s}
for brief review of dual-process theory in reasoning with the list of
necessary references).In what follows we are interested only in human
reasoning where two primary dual systems of interest can be specified. One
of these systems we will call it below as deductive reasoning system (DRS)
is rational, sequental and consistent but acts relatively slow while the
other - we will call it further as heuristic reasoning system (HRS) is
intuitive, fast, automatic, but to a large extent influenced by emotions and
last unconsious experience.Numerous researches and experiments conclusively
proved that there is hidden interaction between these two cognitive\ systems
such that a reasoning subject is not aware of this.The belief-bias effect is
the most striking manifestation of such interaction. Roughly speaking the
belief-bias effect is the innate tendency of reasoning subjects to be more
likely to accept reasons ang arguments if they find them believable and to
pay less attention of their logical validity. The main goal of present paper
is based on quantum ideas of complementarity and dual-process theory in
human reasoning to describe the belief-bias effect phenomenologically by
purely logical tools. To this end we will use also the simplified version of
discrete-continuous logic that was formulated earlier in author preprint 
\cite{5s}.

The remainder of the paper is organized as follows. In chapter 2 we briefly
remind basic facts relating to discrete-continuous logic (DCL) that are
necessary for the understanding of the present paper.The main contribution
of this chapter is the interpretation of the general propositions in DCL as
the integral mental structures that consist both of logical and heuristic
constituents. Under such interpretation these two constituents of the
proposition can be considered as complementary to each other exactly like
two noncommuting observables in quantum mechanics.In chapter 3 we state the
uncertainty relation that just reflects the complementary nature of such
concepts as logical rigour and the heuristic grasp. And finally in chapter 4
using only logical tools we give the phenomenological explanation \ of the
belief-bias effect. Now let us go to the details.

\section{Preliminiries}

In this part we briefly remind for the reader convenience the necessary
facts relating to the discrete -continuous logic that were outlined more
detail in author preprint \cite{5s}. So, we will consider as the primary
objects of our study the set of general propositions (GP) -$\left\{
A_{j}\right\} _{\text{ }}$that may be represented by $2\times 2$ positive
definite matrices with unit trace of the following form:%
\begin{equation}
A_{j}=%
\begin{pmatrix}
p_{j} & i\alpha _{j} \\ 
-i\alpha _{j} & 1-p_{j}%
\end{pmatrix}%
,  \label{f1}
\end{equation}%
(where $i$ is imaginary unit). In this case the negation of such proposition
- ($not$ $A_{j}$) may be defined as $(not$ $A)=%
\begin{pmatrix}
1-p_{j} & -i\alpha _{j} \\ 
i\alpha _{j} & p_{j}%
\end{pmatrix}%
$. It turns out that in addition to negation another but already two place
operation -$\bigtriangleup $ (which is the analogue of strong disjunction in
ordinary Boolean logic) can be introduced in DCL according to the next
definition:

Let $A=%
\begin{pmatrix}
p & i\alpha  \\ 
-i\alpha  & 1-p%
\end{pmatrix}%
$ and $B=%
\begin{pmatrix}
q & i\beta  \\ 
-i\beta  & 1-q%
\end{pmatrix}%
$ then%
\begin{equation}
(A\bigtriangleup B)=%
\begin{pmatrix}
R & i\gamma  \\ 
-i\gamma  & 1-R%
\end{pmatrix}%
,  \label{f2}
\end{equation}%
where $R=p+q-2pq+2\alpha \beta $, $\gamma =\alpha \left( 1-2q\right) +\beta
\left( 1-2p\right) $. Comparing representation Eq. (\ref{f1}) with standard
form of density matrix of the mixed state of two-level quantum system that
looks as $\rho =%
\begin{pmatrix}
\frac{1+P_{z}}{2} & \frac{P_{x}-iP_{y}}{2} \\ 
\frac{P_{x}+iP_{y}}{2} & \frac{1-P_{z}}{2}%
\end{pmatrix}%
$ (where $P=(P_{x},P_{y},P_{z})$ is the Bloch vector of the state ) we see
that GP may be represented by the similar way but in this case $x-$
component of the Bloch vector is equal to zero. In the rest of the paper we
will use such reduced Bloch representation for the arbitrary proposition $A,$
that is: $\ A=%
\begin{pmatrix}
\frac{1+P_{z}}{2} & \frac{-iP_{y}}{2} \\ 
\frac{iP_{y}}{2} & \frac{1-P_{z}}{2}%
\end{pmatrix}%
$ with $P=(P_{y},P_{z})$. In this case it is convenient to introduce the
complex vector $P=P_{z}-iP_{y}$ which we call further as representating
vector (RV) of proposition $A$. It is easy to verify directly that the RV of
proposition $(notA)$ is equal to $(-P)$ and RV of proposition $%
(A\bigtriangleup B)$ is equal to $(-PQ)$ (where $Q$ is RV of $B$ ).Note also
the useful relation connecting negation with operation $\bigtriangleup $:%
\begin{equation}
not(A\bigtriangleup B)\equiv (A\overline{\bigtriangleup }B)=(notA)%
\bigtriangleup B=A\bigtriangleup (notB).  \label{f3}
\end{equation}%
It should be noted that unlike of ordinary Boolean logic in DCL it is
possible to define the whole one-parameter group of continuous logical
operations (logical rotations of propositions in the plane $P_{y}-P_{z}$)
according to the following rule: if proposition $A$ has the RV -$P$ then
rotated at an angle $\Phi $ proposition $A^{%
{{}^1}%
}$ has RV - $P%
{{}^1}%
$ with components:%
\begin{eqnarray}
P_{y}^{%
{{}^1}%
} &=&P_{y}\cos \phi +P_{z}\sin \Phi   \notag \\
P_{z}^{%
{{}^1}%
} &=&P_{z}\cos \Phi -P_{y}\sin \Phi   \label{4f}
\end{eqnarray}%
It is easy to see that the negation of any proposition coincides with
logical rotation of it at an angle $\pi $ and in addition that if one
rotates the GP $A$ at an angle $\Phi _{1}$ and the other proposition $B$ at
\ an angle $\Phi _{2}$ then the proposition $(A\bigtriangleup B)$ will be
rotated at an angle $\Phi _{1}+\Phi _{2}$. Thus all logical operations in
DCL obtain quite clear geometric meaning. Now after describing the syntax of
DCL we can pass to the more difficult task: clarification of its semantics
that is the interpretation both the meaning of general propositions and
logical operations with them. It should be noted that interpretation that we
are going to propose here is not the only possible but it is appropriate for
our ultimate goal namely to explain the beliefe-bias effect in human
reasoning from pure logical point of view. So, as before we will assume that
diagonal elements of representing matrix for arbitrary GP describes degree
of its logical validity (from DRS point of view) while its nondiagonal
elements we will interpret as the believability of the same proposition
inspired by the heuristic reasoning system (HRS). This interpretation can be
expressed more precisely as follows. Let us introduce two projection
operators:$P_{1}$ and $P_{2}$ $\left( P_{1}^{2}=P_{1}\text{, }%
P_{2}^{2}=P_{2}\right) $ according to the definition: $P_{1}=\frac{1+\sigma
_{z}}{2}$ and $P_{2}=\frac{1+\sigma _{y}}{2}$. It is easy to see that
average values of these operators in the state whose density matrix
coincides with representating matrix of proposition $A=%
\begin{pmatrix}
p & i\alpha  \\ 
-i\alpha  & 1-p%
\end{pmatrix}%
$ give us the propabilities of its logical plausability $p_{t}$ and its
believability $p_{b\text{ }}$respectively.Thus we obtain 
\begin{eqnarray}
p_{t} &=&\left\langle P_{1}\right\rangle =Sp(P_{1}A)=p  \notag \\
&&\text{and}  \label{5f} \\
p_{c} &=&\left\langle P_{2}\right\rangle =Sp(P_{2}A)=\frac{1-2\alpha }{2} 
\notag
\end{eqnarray}%
In connection with above interpretation we want \ to point out two important
marginal GP: 1) $T=%
\begin{pmatrix}
1 & 0 \\ 
0 & 0%
\end{pmatrix}%
$-true proposition, and 2) $B=%
\begin{pmatrix}
\frac{1}{2} & -\frac{i}{2} \\ 
\frac{i}{2} & \frac{1}{2}%
\end{pmatrix}%
$ - highest possible believable proposition and their negations: $F=(notT)-$%
false proposition and $U=(notB)$-unbelievable proposition. Note in addition
that noncommutativity of operators $P_{1}$and $P_{2}$ implies that main
predicates of arbitrary GP (plausibility and belief) may be considered as
complementary (in the sence of quantum theory) aspects of the same
proposition. This important fact implies specific uncertainty relation for
the observables $P_{1}$and $P_{2}$ connected with any GP. The simple
derivation of these relation is the subject of the next section of the
presenr paper.

\section{The Uncertainty Relation between predicates plausibility and
believability in DCL.}

To derive the required uncertainty relation it is convenient to represent
any GP $A$ in the Bloch form: $A=%
\begin{pmatrix}
\frac{1+P_{z}}{2} & \frac{-iP_{y}}{2} \\ 
\frac{iP_{y}}{2} & \frac{1-P_{z}}{2}%
\end{pmatrix}%
.$According to definition the uncertainty of logical truth for the
proposition $A$ can be written with the help of operator $P_{1}=\frac{%
1+\sigma _{z}}{2}$ as: $\bigtriangleup p_{t}^{2}\equiv \overline{\left( 
\frac{1+\sigma _{z}}{2}-\overline{\frac{1+\sigma _{z}}{2}}\right) ^{2}}=%
\frac{1}{4}\left( 1-\overline{\sigma _{z}}^{2}\right) =\frac{1-P_{z}^{2}}{4}$%
.In the similar manner the uncertainty of believability of the same
proposition is equal to : $\bigtriangleup p_{c}=\frac{1}{4}\left( 1-%
\overline{\sigma _{y}}^{2}\right) =\frac{1}{4}\left( 1-P_{y}^{2}\right) .$By
adding these two expressions we obtain: $\bigtriangleup p_{A}^{2}\equiv
\bigtriangleup p_{t}^{2}+\bigtriangleup p_{c}^{2}=\frac{1}{4}\left(
2-P_{y}^{2}-P_{z}^{2}\right) .$Finally taking into accout that $%
P_{y}^{2}+P_{z}^{2}\leqslant 1$ we get the desired relations:%
\begin{equation}
\frac{1}{4}\leqslant \bigtriangleup p_{A}^{2}\leqslant \frac{1}{2}.
\label{6f}
\end{equation}%
The notable fact should be mentioned here:if one takes two propositions $A$
and $B$ with RV $P$ and $Q$ respectively then according above calculation
one can write two equations 1) $\bigtriangleup p_{A}^{2}=\frac{\left(
2-P^{2}\right) }{4}$ and 2)$\bigtriangleup p_{B}^{2}=\frac{\left(
2-Q^{2}\right) }{4}.$

On the other hand as we marked earlier the proposition $\left(
A\bigtriangleup B\right) $ has RV $(-PQ)$ and hence its uncertainty is equal
to $\bigtriangleup $ $p_{(A\bigtriangleup B)}^{2}=\frac{\left(
2-P^{2}Q^{2}\right) }{4}$ .As long as $P^{2}Q^{2}\leqq P^{2},Q^{2}$ one can
conclude $\ \ $that $\bigtriangleup p_{(A\bigtriangleup B)}^{2}\geqslant
(\bigtriangleup p_{A}^{2},\bigtriangleup p_{B}^{2})$ and hence as a final
result of logical operation $\bigtriangleup $ the ending uncertainty of
proposition can only increases.We would like to hope that properly organized
experiments with specially selected reasoning tasks will be able to confirm
(or may be disprove) the proposed uncertainty relations $\left( 6\right) $%
.Now we come back to the main goal of present paper: the explanation of the
belief-bias effect in human reasoning.

\section{Many valued probabilistic logic and the Belief-Bias effect.}

In this part we will try to describe (phenomenologically) the belief-bias
effect in human reasoning by purely logical tools.For this purpose it is
convenient to use some version of probabilistic many-valued logic that in
some sense can be considered as simplified version of original DCL.Really if
in original version of DCL we restrict ourselves only by discrete set of
logical rotations with angles: $0,\frac{2\pi }{N}...\frac{2\pi }{N}\left(
N-1\right) $ we obtain the closed logic with N marginal propositions which
possess representing matrices: $A_{0},A_{1}....A_{N-1}$ (where $A_{0}=%
\begin{pmatrix}
1 & 0 \\ 
0 & 0%
\end{pmatrix}%
,...$ $A_{k}=%
\begin{pmatrix}
\frac{1+\cos \frac{2\pi }{N}}{2} & -\frac{i}{2}\sin \frac{2\pi }{N} \\ 
\frac{i}{2}\sin \frac{2\pi }{N} & \frac{1-\cos \frac{2\pi }{N}}{2}%
\end{pmatrix}%
(k=1..N-1)$)$.$ In the case when we are not interested in the "quantum
correlations" between these marginal propositions we can consider them as
approximately independent quantities and associate with these propositions
the logical basis consisting of N distinct logical alternatives.Acting in
this manner one can pass from original DCL to standard many- valued
probabilistic Boolean logic. After this remark we will examine further four
valued probabilistic logic every proposition of which can be represented as $%
4\times 4$ diagonal matrix : $A=diag\left( P_{1},P_{2},P_{3},P_{4}\right) $%
.Here we mean that the space of these propositions is a tensor product of
two spaces with $2\times 2$ \ diagonal matrices, that is:%
\begin{equation}
A=\sum\limits_{i}a_{i}T_{i}\otimes B_{i},  \label{7f}
\end{equation}%
where $T_{i}=%
\begin{pmatrix}
p_{i} & 0 \\ 
0 & 1-p_{i}%
\end{pmatrix}%
,$ $B_{i}=%
\begin{pmatrix}
q_{i} & 0 \\ 
0 & 1-q_{i}%
\end{pmatrix}%
$and $\sum\limits_{i}a_{i}=1$. In addition we assume that matrices $T_{i}$
in the decomposition Eq. (\ref{7f}) are associated with the activity of
deductive cognitive subsystem (DRS),while matrices $B_{i}$ are connected
with its heuristic subsystem (HRS).Thus the basis of this logic consists of
four propositions:$1)$ truth-believable $TB=diag(1,0,0,0),2)$
truth-unbelievable $TU=diag(0,1,0,0),3)$false-believable $FB=diag(0,0,1,0)$
and $4)$false- unbelievable $FU=diag(0,0,0,1).$Our next step is to define
basic logical operations that can be implemented with such propositions. The
interpetation that we have adopted above implies that the negation of
proposition $A$ must be defined as $\left( notA\right) =diag\left(
P_{4},P_{3},P_{2},P_{1}\right) $. The certain dilemma arises however when we
want to define the conjunction of two propositions $A=diag\left(
P_{1},P_{2},P_{3},P_{4}\right) $ and $B=diag\left(
Q_{1},Q_{2},Q_{3},Q_{4}\right) $. We have proposed here the following
definition:%
\begin{equation}
C\equiv \left( AandB\right) =diag\left( C_{1},C_{2},C_{3},C_{4}\right) ,
\label{8f}
\end{equation}%
where $C_{1}=P_{1}\left( Q_{1}+Q_{2}\right) +P_{2}Q_{1}$, $C_{2}=P_{2}Q_{2}$%
, $C_{3}=P_{1}\left( Q_{3}+Q_{4}\right) +P_{2}Q_{3}+P_{3}+P_{4}\left(
Q_{1}+Q_{3}\right) $, $C_{4}=P_{2}Q_{4}+P_{4}\left( Q_{2}+Q_{4}\right) $.
This definition of conjuction namely Eq. (\ref{8f}) certainly needs to be
explained.First of all we note that definition Eq. (\ref{8f}) satisfies to
the necessary symmetry condition : $\left( AandB\right) =\left( BandA\right) 
$as it should be.In addition if one takes the projection of conjunction Eq. (%
\ref{8f}) in DRS (first reasoning subsystem) the result is:$\left(
AandB\right) _{1}=%
\begin{pmatrix}
pq & 0 \\ 
0 & 1-pq%
\end{pmatrix}%
\equiv \left( A_{1}andB_{1}\right) $ where $p=P_{1}+P_{2}$ and $q=Q_{1}+Q_{2}
$. This result obviously consistent with definition of conjunction in
ordinary probabilistic Boolean logic. On the other hand if one takes the
projection of Eq. (\ref{8f}) in HRS (second reasoning subsystem) the
obtained result reads as:%
\begin{equation}
\left( AandB\right) _{2}=%
\begin{pmatrix}
1-\left( P_{2}+P_{4}\right) \left( Q_{2}+Q_{4}\right)  &  \\ 
\left( P_{2}+P_{4}\right) \left( Q_{2}+Q_{4}\right)  & 
\end{pmatrix}%
.  \label{9f}
\end{equation}%
We see that conjunction in heuristic system differs from standatd logical
conjunction .In our opinion this distinction explicitly reflects (from
phenomenological point of view) the essential difference existing between
two reasoning systems when they operate jointly. In particular the
definition Eq. (\ref{9f}) implies for two basic marginal propositions in
second reasoning subsystem: $B=%
\begin{pmatrix}
1 & 0 \\ 
0 & 0%
\end{pmatrix}%
$ responding to the statement of unconditional belief and $D=%
\begin{pmatrix}
0 & 0 \\ 
0 & 1%
\end{pmatrix}%
$ -which is the most doubtful statement, the next conjunction relations:$%
\left( BandB\right) =B$, $\left( BandD\right) =\left( DandB\right) =B$ and $%
\left( DandD\right) =D$.Thus we obtain that the unconditional belief when it
conflicts with certain doubtful one always overcomes it. Now if one takes
the expression Eq. (\ref{8f}) for granted then he (she)can define another
logical operations (in particular implication that we especially interested
in ) without any obstacles.To this end one should be guided by two relations
of ordinary logic which as we assume continue to be valid in our case as
well: 1) $\left( AorB\right) =not\left[ \left( notA\right) and\left(
notB\right) \right] $ and 2) $(A\Longrightarrow B)=\left( notA\right) orB$
.Acting in this manner we obtain for the implication $(A\Longrightarrow B)$
\ the required relation:%
\begin{equation}
I\equiv (A\Longrightarrow B)=diag\left( I_{1},I_{2},I_{3},I_{4}\right) ,
\label{10f}
\end{equation}%
where $I_{1}=p_{4}\left( q_{1}+q_{3}\right) +p_{2}q_{1}$, $%
I_{2}=p_{3}+q_{2}\left( 1-p_{3}\right) +p_{1}q_{1}+p_{4}q_{4}$, $%
I_{3}=p_{2}q_{3}$, $I_{4}=p_{1}\left( q_{3}+q_{4}\right) +p_{2}q_{4}$. The
expression Eq. (\ref{10f}) for the implication of two probabilistic
propositions in four- valued logic is the foundation for our following
explanation of bias-belief effect.Note that here we are going to demonstrate
only the simplest case of the application of the approach proposed. The
detail quantitative analysis of numerous possible situations connected with
the interaction between DRS and HRS will be realized by us at length in
separate publication. So, let us take the proposition $B$- (consequent of
the implication) in the form: $B=diag\left( 1,0,0,0\right) $, that means
that consequent is both true and believable proposition. Then the expression
Eq. (\ref{10f}) implies that matrix $\left( A\Longrightarrow B\right) $ has
the form:%
\begin{equation}
(A\Longrightarrow B)=diag\left( p_{2}+p_{4},p_{1}+p_{3},0,0\right) ,
\label{f11}
\end{equation}%
and hence its projections in DRS (1) and HRS (2) systems are respectively : $%
\left( A\Longrightarrow B\right) _{1}=%
\begin{pmatrix}
1 &  \\ 
& 0%
\end{pmatrix}%
$ , and $\left( A\Longrightarrow B\right) _{2}=%
\begin{pmatrix}
p_{2}+p_{4} &  \\ 
& p_{1}+p_{3}%
\end{pmatrix}%
$

On the other hand if one choose the consequent $B$ in the form $B=diag\left(
0,1,0,0\right) $ that means that consiquent $B$ is true but unbelievable
proposition then according to expression $\left( 10\right) $ one obtain for
the implication $\left( A\Longrightarrow B\right) $ the relation:%
\begin{equation}
\left( A\Longrightarrow B\right) =diag\left( 0,1,0,0\right) ,  \label{12f}
\end{equation}%
and hence the projections of this proposition in two cognitive systems are:$%
\left( A\Longrightarrow B\right) _{1}=%
\begin{pmatrix}
1 &  \\ 
& 0%
\end{pmatrix}%
$ and $\left( A\Longrightarrow B\right) _{2}=%
\begin{pmatrix}
0 &  \\ 
& 1%
\end{pmatrix}%
.$

Now if we make the natural assumption that after the first ( unconscious)
stage of reasoning, when two cognitive systems operate jointly, at the
second stage the conscious evaluation of the validity of a conclusion $V$
occurs in accordance with the simple rule:%
\begin{equation}
V=aP_{t}+\left( 1-a\right) P_{b},  \label{11f}
\end{equation}%
(where $a$ $(0\leqslant a\leqslant 1)$ is certain number coefficient
depending on age,intellect,training of the subject and possibly some other
factors).Note that this assumption in fact coincides with similar rule which
was used in the paper \cite{4s}. Now returning to the above example of
interest we result in that the magnitude of the bias-belief effect $V$ can
be evaluate quantitatively as $V\equiv V_{1}-V_{2}=\left( 1-a\right) \left(
p_{2}+p_{4}\right) $.We believe that although the value of coefficient $a$
is unknown in advance nevertheless the validity of the Eq. (\ref{11f}) can
be explicitly verified in seria of properly organized psychological
experiments with various subjects using the identical cognitive tasks .

In conclusion of our study let us formulate once more the central results of
the present paper:

1)We introduced the novel version of DCL with both discrete and continuous
logical operations between generalized propositions .

2)We proposed the concrete interpretation of propositions in DCL as integral
mental structures that include both logical and heuristic constituents.

3)We stated the specific uncertainty relation between logic rigour and
heuritic grasp that reflect complementary aspects of human reasoning process.

4)We proposed phenomenological model of human reasoning based on simplified
version of DCL and demonstrated that it is able to explain belief-bias
effect qualitatively and possibly \ \ quantitatively as well.

All these conclusions we hope to discuss more detail in our further
publications.

\end{document}